\documentclass[aps,prd,widetext,showpacs,preprintnumbers,twocolumn,superscriptaddress,nofootinbib]{revtex4}

\usepackage[dvipdfmx]{graphicx}       
\usepackage{amssymb}
\usepackage{amsmath}
\usepackage{dcolumn}
\usepackage{bm}
\usepackage{natbib}
\usepackage{multirow}
\usepackage{subfigure}
\newcommand{\non}{\nonumber}  
\newcommand{\lmk}{\left(}  \newcommand{\rmk}{\right)}

\newcommand{\vect}[1]{\mbox{\boldmath${#1}$}}

\begin{document}

\preprint{DESY\ 15-207}

\title{Possible resolution of the domain wall problem in the NMSSM}

\author{Anupam Mazumdar}
\affiliation{Consortium for Fundamental Physics, Lancaster University, Lancaster, LA1 4YB, UK}
\affiliation{IPPP, Durham University, Durham, DH1 3LE, UK}

\author{Ken'ichi Saikawa}
\affiliation{Department of Physics, Tokyo Institute of Technology,
2-12-1 Ookayama, Meguro-ku, Tokyo 152-8551, Japan}
\affiliation{Deutsches Elektronen-Synchrotron DESY,
Notkestrasse 85, 22607 Hamburg, Germany}

\author{Masahide Yamaguchi}
\affiliation{Department of Physics, Tokyo Institute of Technology,
2-12-1 Ookayama, Meguro-ku, Tokyo 152-8551, Japan}

\author{Jun'ichi Yokoyama}
\affiliation{Research Center for the Early Universe (RESCEU),
Graduate School of Science, The University of Tokyo, Tokyo 113-0033,
Japan}
\affiliation{Department of Physics, Graduate School of Science, The
University of Tokyo, Tokyo 113-0033, Japan}
\affiliation{Kavli Institute for the Physics and Mathematics of the Universe (Kavli IPMU),
UTIAS, WPI, The University of Tokyo, Kashiwa, Chiba 277-8568, Japan}

\date{\today}

\begin{abstract}
We discuss a possibility that the domain wall problem in the next-to-minimal
supersymmetric standard model is alleviated without introducing a small explicit $Z_3$ breaking term
by analyzing the evolution of the singlet scalar field within an inflationary paradigm.
The singlet scalar field which explains the
$\mu$-term tracks a time-varying minimum of the effective potential
after inflation and slowly rolls down to its global minimum if there
exist sufficiently large
negative Hubble-induced corrections on the effective potential for the singlet field, which arise through
supergravity.  As a consequence, the whole Universe is
confined within a single domain during and after inflation, which
prevents the formation of domain walls. This will further constrain the history of the early Universe 
along with the Higgs-singlet coupling.
\end{abstract}


\maketitle

\section{Introduction}

Supersymmetry (SUSY) is  the
most popular and plausible paradigm to resolve the hierarchy problem in nature
between the grand unification scale and the electroweak scale; for a 
review, see~\cite{Nilles:1983ge,Martin:1997ns}. The SUSY is also highly
attractive from cosmological viewpoints~\cite{Enqvist:2003gh}.
It provides an appropriate candidate for cold dark matter in terms  of the lightest SUSY
particle (LSP)~\cite{Jungman:1995df}. 
It is also useful to preserve the flatness of the inflaton potential against
radiative corrections. Indeed one can find a number of candidates for
the inflaton in models with SUSY or supergravity (see, {\it e.g.}, 
\cite{Lyth:1998xn,Mazumdar:2010sa,Yamaguchi:2011kg} for reviews), making use of a
gauge-singlet multiplet  \cite{Murayama:1992ua,Murayama:1993xu,Kadota:2005mt}, {\it gauge invariant} flat
directions
\cite{Allahverdi:2006iq,Allahverdi:2006we}
(see also \cite{Kamada:2009hy}), or Higgs fields having a nonminimal
scalar-curvature coupling~\cite{Ferrara:2010yw,Ferrara:2010in}. 
Finally, an efficient mechanism of baryogenesis has been proposed in
SUSY, making use of flat directions \cite{Affleck:1984fy}.

The minimal version of the SUSY standard model (SM), also known as MSSM, 
contains one dimensionful parameter in the superpotential, namely,
the $\mu$-term, i.e., $\mu H_uH_d$, where the SU(2) doublets
$H_u $ and $ H_d $ 
yield masses to uplike  and downlike quarks as they acquire
vacuum expectation values (VEVs), respectively.
It is desirable that the origin of such a dimensionful parameter
as well as its magnitude, $\mu \sim {\cal O}({\rm TeV})$, 
be explained by a more fundamental theory.
Along this line, it has been proposed to  extend MSSM to incorporate  
an additional singlet chiral superfield $S$, which dynamically generates the
$\mu$-term~\cite{Nilles-1,Kim:1983dt}. 
This is also known as next-to-MSSM (NMSSM); for a review, see~\cite{Ellwanger:2009dp}. 

Terms dependent on an {\it absolute gauge singlet}, $S$,
in the renormalizable superpotential for the NMSSM read 
\begin{equation}
W = \lambda S H_u H_d + \frac{\kappa}{3}S^3 + W_{\rm MSSM}, \label{W_NMSSM}
\end{equation}
where $W_{\rm MSSM}$ represents the usual Yukawa interactions between
Higgs doublets and quarks/leptons in the MSSM, and $\lambda$ and
$\kappa$ are dimensionless couplings. In this model, typically a
discrete $Z_3$ symmetry is imposed under which all chiral superfields
$\Phi$ transform as $\Phi\to e^{2\pi i/3}\Phi$.  Such a symmetry
guarantees the absence of terms like $\propto S$ and $\propto S^2$ as
well as the $\mu$-term in the MSSM.

The VEV of $\langle S \rangle $, however, will spontaneously break the $Z_3$ symmetry, such that
$\mu = \lambda \langle S\rangle \sim
\mathcal{O}(10^2\mathchar`-10^3)\mathrm{GeV}$ 
in order to explain the low-scale SUSY spectrum and the observed Higgs
mass. This symmetry breaking
also poses an intriguing problem and a challenge, which leads to the formation of
domain walls in the Universe. 
Although domain walls with tiny energy scale may yield some interesting
cosmological consequences \cite{Frieman:1991tu} including mild
acceleration of cosmic expansion \cite{Brandenberger:2003ge},
in the present case, their energy scale is
so high that if they persist in the late Universe, they simply cause
cosmological disasters overdominating the energy density of the
Universe \cite{Zeldovich:1974uw}.

In the literature, in order to evade this problem,
it is usually assumed that the $Z_3$ symmetry is an
accidental symmetry and there exists some explicit symmetry breaking
term, which leads to the collapse of domain walls at late
times.\footnote{For instance, the small explicit $Z_3$ breaking term can
be obtained by imposing a discrete subgroup of $U(1)_R$
symmetry~\cite{Panagiotakopoulos:1998yw}.
The large explicit $Z_3$ breaking term can also be generated in the context of
superconformal embedding of NMSSM into supergravity~\cite{Ferrara:2010in}.}
The purpose of the present paper is to provide an alternative explanation 
to avoid the domain wall problem in the context of inflationary cosmology
rather than an explicit symmetry breaking term employed so far in the literature. 
We will see that the formation of domain walls can be alleviated if there exist sufficiently large
negative supergravity corrections proportional to the Hubble parameter
in the effective potential for the singlet scalar field during and after inflation.\footnote{A similar scenario was considered in Ref.~\cite{McDonald:1997vy},
in which a new gauge singlet field is introduced in addition to the NMSSM field content and this new scalar field acquires a large Hubble-induced mass.
Instead of introducing such an extra symmetry breaking field,
in this paper we discuss a possibility that the domain wall problem is avoided solely due to the dynamics of the singlet scalar field $S$ in NMSSM.}
Since the singlet field tracks a time-varying minimum of the effective potential after inflation, we call this scenario the {\it tracking mechanism}.

The outline of the paper is as follows: In Sec.~\ref{sec2}, we discuss
the domain wall formation in the inflationary context. Then possible
effects caused by supergravity corrections are described in
Sec.~\ref{sec3}. In Sec.~\ref{sec4}, we analyze the evolution of
the singlet field with a negative Hubble-induced mass and derive the
conditions for the tracking mechanism to work. Initial conditions for
the singlet field during inflation are also discussed in
Sec.~\ref{sec5}. In Sec.~\ref{sec6}, we take account of finite
temperature corrections for the evolution of the singlet field after
inflation and obtain some conditions on the Higgs-singlet coupling and
the reheating temperature in order to avoid the domain wall formation.
The consequences for cosmology in the NMSSM are briefly discussed in
Sec.~\ref{sec7}. Finally, Sec.~\ref{sec8} is devoted to the conclusion and discussions.

\section{NMSSM domain walls and inflation\label{sec2}}

In order to address this issue, let us first note that from
$\mu=\lambda\langle S\rangle\sim
\mathcal{O}(10^2\mathchar`-10^3)~\mathrm{GeV}$, we see that $\langle
S\rangle$ can be much larger than the weak scale if the coupling
$\lambda$ is sufficiently small.  In this case, the scalar potential can
be written as $$V \simeq \kappa^2|S|^4 + m_S^2|S|^2 +
\left(\frac{\kappa}{3} A_{\kappa}S^3 + \mathrm{h.c.}\right),$$ where
$m_S^2$ and $A_{\kappa}$ are the soft SUSY breaking mass
parameters.\footnote{In principle, superpotential terms such as
$W\subset {\cal O}(1) S^{n}/M_{\rm Pl}^{n-3}$ would also appear, where
$n= 6,~9,\cdots $, and $M_{\rm Pl}\simeq 2.4\times 10^{18}$~GeV is the
reduced Planck mass.  In this paper we are ignoring these higher order
contributions to the superpotential and the potential.}  From the above
form of the potential, we can estimate the VEV of $S$ at the global
minimum as
\begin{equation}
\langle S \rangle_{\rm global}\simeq -\frac{A_{\kappa}}{4\kappa}\left(1+\sqrt{1-\frac{8m_S^2}{A_{\kappa}^2}}\right). \label{S_VEV_global}
\end{equation}
Note that the magnitude of $A_{\kappa}$ must be slightly larger than that of $m_S$ in order to guarantee the existence of the global minimum with $\langle S\rangle\ne 0$~\cite{Ellwanger:1996gw}:
\begin{equation}
A_{\kappa}^2/m_S^2 \gtrsim \mathcal{O}(10). \label{condition_A_term}
\end{equation}
Furthermore, $\lambda$ and $\kappa$ cannot have a large hierarchy if the
magnitudes of all dimensionful parameters $\mu$, $m_S$, and $A_{\kappa}$
are close to the weak scale. In this paper, we  assume $\lambda
\simeq \kappa$ for simplicity.

Before discussing the resolution of the problem, let us describe how
likely the formation of domain walls occurs in the inflationary context.
For simplicity, we assume that inflation is driven by a potential 
energy of some
scalar field, called inflaton, and that the inflaton sector does not
embody the $S$ superfield, or as a matter of fact other MSSM
superfields. A naive expectation is that the domain wall formation is
avoidable if the thermal corrections to the effective potential of the
singlet scalar remain irrelevant after inflation such that the $Z_3$
symmetry is never restored, but this simple observation turns out to be
insufficient when we carefully consider the evolution of the singlet
scalar during and after inflation due to the reasons discussed below.

During inflation, quantum mechanically induced vacuum fluctuations
significantly displace any light scalar field whose effective mass is
smaller than the Hubble parameter $H_{\rm inf}$ because it obtains
quantum fluctuations of order $H_{\rm inf}/2\pi$ within each Hubble
time~\cite{Vilenkin:1982wt,Linde:1982uu,Starobinsky:1982ee}. After a
sufficiently large number of $e$-folds of inflation, a free real scalar
field $\phi$ with mass $m_\phi$ acquires long-wave fluctuations with
dispersion
\begin{equation}
 \langle \phi^2 \rangle=\frac{3H_{\rm inf}^4}{8\pi^2 m_\phi^2}.
\end{equation}
The dispersion of a scalar field with a more complicated  potential 
can be calculated by 
the stochastic inflation method \cite{Starobinsky:1994bd} 
but the typical field amplitude can be estimated by simply
requiring that the typical value of its potential energy density takes
a value $\mathcal{O}(H_{\rm inf}^4)$.

This fact implies that the VEV of the $S$ field during inflation is
generically different from the value $\langle S\rangle_{\rm
global}$,~Eq.~\eqref{S_VEV_global}, at the global minimum of the
low-energy effective potential because the bare mass of $S$ field
is $m_S\sim {\cal O}({\rm TeV}) \ll H_{\rm inf}$. After inflation, the
$S$ field starts to oscillate, reducing its amplitude with time. It eventually falls into one of the global minima, whose phase is related by the $Z_3$ transformation to that of other minima.

Note, however, that the final value of the phase of $\langle S\rangle$ can
differ at each spatial point because of the existence of the field
fluctuations $\delta S(x)$. These field variations originate from
quantum fluctuations during inflation, and they may be enhanced once the
$S$ field starts to oscillate due to the parametric resonance effect; see~\cite{Kofman:1994rk}.
The enhancement occurs both in the radial and angular
direction of the complex scalar $S$, since the  $A_{\kappa}$
term mixes them.  As a result, different domains are created within the
horizon scale, and domain walls are formed around their boundary.  
The above discussion suggests that the formation of domain walls is
almost inevitable unless the $S$ field is stabilized at the minimum of
the effective potential during inflation.

\section{Supergravity correction and the evolution of $S$\label{sec3}}

In this paper, we shall argue
that the resolution of this problem is also achieved within the
inflationary paradigm (For a recent review of inflation, see, {\it e.g.},~\cite{Sato:2015dga}.)
The key ingredient is an effective mass of the
form $-cH^2$ for the $S$ field, where $H$ is the Hubble parameter at a
given time.  Such a mass term generically arises via a Planck-suppressed
interaction in the framework of supergravity~\cite{Dine:1995uk}; see 
also~\cite{Lyth:1998xn,Enqvist:2003gh,Mazumdar:2010sa,Yamaguchi:2011kg}.  Since
the value of the coefficient $c$ depends on the details of the K\"ahler
terms~\cite{Choudhury:2014sxa},  it is fair to treat it as a free parameter to keep the discussion
model independent. According to its value, we may consider the following three
possibilities:

\begin{itemize}

\item \underline{If there exists no Hubble-induced mass ($c=0$)}: In
this case, $S$ is expected to take a large value $S\sim H_{\rm inf}/\sqrt{\kappa}$
during inflation due to the accumulation of long wave quantum
fluctuations.  It starts oscillation as the Hubble parameter gets
smaller than the effective mass of $S$ after inflation.  Then even in case a motion
along the angular direction was suppressed initially, as it
crossed the origin, it would start the angular motion and the phase of
the scalar field would have a scattered distribution due to the
$Z_3$ symmetry of the potential.  As a result, formation of domain
walls is inevitable.

\item \underline{If the Hubble-induced mass is positive ($c<0$)}: In
this case, during inflation the $S$ field is stabilized at the
origin. After inflation, it rolls down to the global minimum from the
origin when the temperature of the Universe becomes sufficiently low
(i.e., the phase transition occurs), and again domain walls are created
through the Kibble mechanism.

\item \underline{If the Hubble-induced mass is negative ($c>0$)}: 
In this case, during inflation the $S$ field takes a value larger 
than $\langle S\rangle_{\rm global}$, very similar to our Fig.~\ref{fig1}. 
As we will see below, there is a possibility to avoid the domain wall
formation in this case. Indeed if $S$ sits on its potential
minimum at each time and adiabatically traces its time evolution 
until the field relaxes to $\langle S\rangle_{\rm global}$,
we may avoid the domain wall formation. 

\end{itemize}

%\if0

\begin{figure}[htbp]
\includegraphics[width=0.45\textwidth]{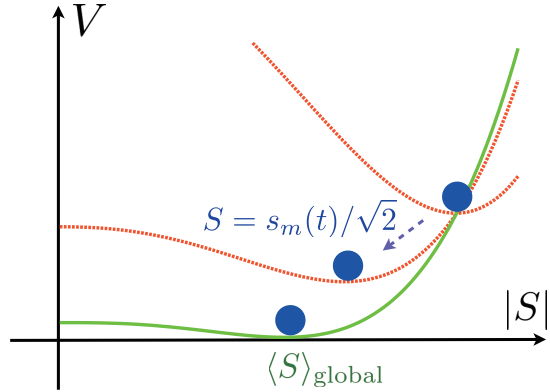}
\caption{Sketch of the tracking mechanism. 
The red dotted lines represent the form of the effective potential with the negative mass term $-\tilde{c}H^2|S|^2$ at early times,
and the green solid line represents that at the present time where the Hubble-induced term becomes irrelevant.
The location of the minimum varies with time, which is indicated by blue circles.}
\label{fig1}
\end{figure}

%\fi

From here onwards, we will only consider the last case with a negative
Hubble-induced mass ($c>0$) to see if the above-mentioned tracking
scenario works.  We can write the effective potential for the $S$ field
during inflation as
\begin{equation}
V = -cH_{\rm inf}^2|S|^2-\left(\frac{\kappa}{3}c'H_{\rm
			  inf}S^3+\mathrm{h.c.}\right) + \kappa^2|S|^4, \label{4}
\end{equation}
where terms dependent on $H_{\rm inf}$ are induced by
the inflaton's potential through the supergravity effect.  
Here we have ignored the soft SUSY breaking contributions, since 
$m_S,A_{\kappa}\ll H_{\rm inf}$, and 
the possible interaction terms with two Higgses, since 
they are stabilized at the origin due to the
large field value of $S$.  The cubic term of the form $\kappa c'HS^3/3$
with $c'$ being another coefficient plays an important role in
stabilizing the phase of $S$ to the minimum of the potential.

Terms dependent on the Hubble parameter $H(t)$ induced by supergravity
effects are present even in the field oscillation regime after inflation:
\begin{equation}
V(S)=-\tilde{c}H^2|S|^2-\left(\frac{\kappa}{3} \tilde{c'}HS^3+\mathrm{h.c.}\right) 
+ \kappa^2|S|^4. \label{5}
\end{equation}
Here the coefficients, $\tilde{c}$ and $\tilde{c'}$, may take somewhat
smaller values than $c$ and $c'$ during inflation, respectively, because
in this regime the kinetic energy is comparable to the potential energy.
Hereafter we assume that all these four parameters are real and
positive, and that ${\cal O}(c) \simeq {\cal O}(\tilde{c})$ and ${\cal O}(c')
\simeq {\cal O}(\tilde{c'})$ for simplicity. It should be noticed that
(the origins of) the phases of the $A$ terms in Eqs.~\eqref{4} and~\eqref{5}
might not be identical if the K\"ahler structures responsible for the
$A$ terms are changed. Since the K\"ahler structures remain unchanged
during inflation and inflaton oscillation due to the energy dominance of
the inflaton field, both phases are expected to coincide. On the other hand, they may
change generally after the inflaton decays and radiation dominates the
energy density of the Universe. Here we assume both phases
coincide for simplicity.

\section{Tracking of the instantaneous minimum for the $S$ field\label{sec4}}

In order for the tracking scenario to work, we must address the
following two issues, apart from thermal effects which will be discussed
later. 
First, it must trace the time evolution of the minimum adiabatically in
the postinflationary Universe when the effective potential is given by 
Eq.~\eqref{5}. 
Second, the initial fluctuations of the singlet field generated during inflation are small 
enough not to create different domains at a later time.
In this section we discuss the former issue. 

In terms of $S\equiv se^{i\theta}/\sqrt{2}~(s \geq 0)$, Eq.~\eqref{5}
is expressed as
\begin{equation}
V(s,\theta)=-\frac{\tilde{c}}{2}H^2s^2
-\frac{\kappa \tilde{c'}}{3\sqrt{2}}Hs^3\cos 3\theta
+ \frac{\kappa^2}{4}s^4.
\end{equation}
The angular direction has one of the minima at $\theta=0$, where the radial
direction feels the potential
\begin{equation}
V(s,0)=-\frac{\tilde{c}}{2}H^2s^2
-\frac{\kappa \tilde{c'}}{3\sqrt{2}}Hs^3
+ \frac{\kappa^2}{4}s^4,
\end{equation}
which is minimized at
\begin{equation}
s=\frac{1}{2\sqrt{2}\kappa}\lmk
 \tilde{c'}+\sqrt{\tilde{c'}^2+8\tilde{c}}\rmk H(t)\equiv
s_{m}(t).
\end{equation}
Neglecting the motion in the phase direction and fixing as $\theta=0$,
we obtain the equation for the radial direction:
\begin{equation}
\ddot{s} + 3H\dot{s} - \tilde{c}H^2s - \frac{1}{\sqrt{2}}\kappa \tilde{c'}Hs^2 + \kappa^2 s^3 = 0.
\end{equation}
By introducing a new variable $\xi=s/s_m$, the above equation can be rewritten as
\begin{align}
&\xi'' + F\xi' +(Ht)^2\xi\nonumber\\
&\times \left[\left(\tilde{c}+\frac{\xi}{8}\left(\tilde{c'}+\sqrt{\tilde{c'}^2+8\tilde{c}}\right)^2\right)(\xi-1)+G\right] = 0, \label{EOM_xi}
\end{align}
where 
\begin{eqnarray}
F  &\equiv & (Ht)\left(2\frac{\dot{H}}{H^2}+3\right) - 1, \nonumber \\
G &\equiv& \frac{\ddot{H}}{H^3} + 3\frac{\dot{H}}{H^2},
\end{eqnarray}
and $\xi'\equiv d\xi/d\ln t=t\dot{\xi}$. The tracking scenario works
only if $\xi\simeq 1$ holds throughout its evolution.

Note that for $a(t)\propto t^p$, where $a(t)$ represents the scale factor of the Universe at a given time, we have $F=3(p-1)$
and $G=(2-3p)/p^2$. Just after inflation the inflaton field oscillates around the minimum of its potential behaving like a matter component, which leads to
$p=2/3$ for the cosmic expansion. Then we have $G=0$, and $\xi=1$ becomes a solution of the field equation [Eq.~\eqref{EOM_xi}].
However, this solution is unstable since the damping is negative, $F=-1<0$.
In other words, if there exists a deviation from $\xi=1$ initially, it grows with time and spoils the tracking mechanism
when the deviation becomes $\mathcal{O}(1)$.

The instability described above can be alleviated if the coefficient
$\tilde{c}$ or $\tilde{c'}$ is sufficiently large.  Note that just
before the end of inflation ($|\dot{H}|/H^2\lesssim 1$) the damping is
positive ($F>0$), and the field variable $\xi$ exponentially converges
into the value determined by setting the bracket in the left-hand
side of Eq. (\ref{EOM_xi}) to zero,
which reads
\begin{equation} 
\xi \simeq 1 - \frac{4G}{\left(\tilde{c'}+\sqrt{\tilde{c'}^2+8\tilde{c}}\right)\sqrt{\tilde{c'}^2+8\tilde{c}}} 
\end{equation}
for $\tilde{c},\tilde{c'}\gg1$. The effects of quantum
fluctuations will be discussed and shown to be negligible in the next
section. Since $G\sim\mathcal{O}(1)$ just before the end of inflation,
this fact implies that at this epoch the value of $\xi$ deviates from
$\xi=1$ by the following quantity:
\begin{equation}
\delta\xi_i\sim\mathcal{O}\left(\tilde{c}^{-1},\ \tilde{c'}^{-2}\right).
\end{equation}

After inflation, the initial deviation $\delta\xi_i$ starts to grow. 
To see how it grows, let us substitute $\xi=1+\delta\xi$ with $\delta\xi\ll 1$ into Eq.~\eqref{EOM_xi}.
Assuming that $F=-1$, $G=0$, and $Ht=2/3$ during the inflaton-oscillation
dominated phase, we have
\begin{equation}
\delta\xi''-\delta\xi'+\frac{1}{9}
 \left(\tilde{c'}+\sqrt{\tilde{c'}^2+8\tilde{c}}\right)
 \sqrt{\tilde{c'}^2+8\tilde{c}}~\delta\xi=0,
\end{equation}
where we neglected the terms of higher order in $\delta\xi$.
The above equation implies that the deviation grows with time as
$\delta\xi \propto t^{1/2}$ for $\tilde{c},\tilde{c'}\gg1$.

In order to guarantee that the deviation from the minimum $s=s_m$ remains small throughout its evolution, we require the following condition:
\begin{equation}
|\delta\xi(t_g)|=\left(\frac{t_g}{t_i}\right)^{\frac{1}{2}}|\delta\xi_i| = \left(\frac{s_{m,\mathrm{inf}}}{s_{m,\mathrm{global}}}\right)^{\frac{1}{2}}|\delta\xi_i|\ll 1, \label{condition_dxi}
\end{equation}
where $t_g$ and $t_i$ represent the time at which the $S$ field reaches the global minimum and that at the end of inflation, respectively.
In the second equality of the above equation, we used the fact that $s_m(t)\propto H(t) \propto 1/t$.
Here $s_{m,\mathrm{inf}}$ corresponds to the value at the end of inflation
\begin{equation}
s_{m,\mathrm{inf}} \equiv \frac{1}{2\sqrt{2}\kappa}\left(c'+\sqrt{c'^2+8c}\right)H_{\rm inf}, \label{s_m_inf}
\end{equation}
and $s_{m,\mathrm{global}}\equiv \sqrt{2}\langle S\rangle_{\rm global}$ corresponds to that at the global minimum [see Eq.~\eqref{S_VEV_global}].
The condition given by Eq.~\eqref{condition_dxi} implies that 
\begin{equation}
c,\tilde{c}\gg \mathcal{O}\left(\frac{H_{\inf}}{|A_{\kappa}|}\right)^{\frac{2}{3}} \quad\mathrm{or}\quad c',\tilde{c'}\gg \mathcal{O}\left(\frac{H_{\inf}}{|A_{\kappa}|}\right)^{\frac{1}{3}}. \label{bound_c}
\end{equation}

We summarize some representative values for the coefficients of the Hubble-induced correction terms in Table~\ref{tab1}.
According to the amplitude of the Hubble parameter during inflation (and hence the energy scale of inflation), 
relatively large coefficients are required. These required values might be much larger than
those derived in Ref.~\cite{Linde:1996cx} in the context of the cosmological moduli problem.
This is because there exist both positive and negative Hubble-induced terms in the model considered in~\cite{Linde:1996cx},
in contrast to the scenario in this paper where there exists the negative Hubble-induced term only.
In the latter case the dragging of the scalar field is not so efficient as we expect in the former case.
Likewise, we expect that the dragging mechanism of Ref.~\cite{Linde:1996cx} does not relax 
the moduli problem if the effective potential possesses the negative Hubble-induced term only.

{\tabcolsep = 2mm
\begin{table}
\begin{center} %\scriptsize
\caption{Lower bounds on the coefficients of the Hubble-induced corrections given by Eq.~\eqref{bound_c} for
some choices of the energy scale of inflation $V_{\rm inf}^{1/4}$.
Here we use the relation $H_{\rm inf}^2=V_{\rm inf}/3M_{\rm Pl}^2$ and the value for the soft parameter $|A_\kappa|=1\mathrm{TeV}$.}
\vspace{3mm}
\begin{tabular}{l c c c}
\hline\hline
$V_{\rm inf}^{1/4}$ & $2\times 10^{12}$GeV & $2\times 10^{15}$GeV  \\
\hline
$H_{\rm inf}$ & $10^6$GeV & $10^{12}$GeV \\
\hline
$c$, $\tilde{c}$ & $\gtrsim 100$ & $\gtrsim 10^6$ \\
$c'$, $\tilde{c'}$ & $\gtrsim 10$ &  $\gtrsim 10^3$ \\
\hline\hline
\label{tab1}
\end{tabular}
\end{center}
\end{table}
}

We also note that at the time $t\lesssim t_g$, the $S$ field does not rotate in the phase direction as long as the
$A_{\kappa}$ term satisfies the condition given by Eq.~\eqref{condition_A_term}.
Then the transition from one vacuum to others related by the $Z_3$ transformations is prohibited
even after the negative Hubble-induced terms become irrelevant.

\section{Initial condition for $S$ field during inflation\label{sec5}}

Let us next study the field configuration during inflation, keeping the above
parameter values in mind. The potential minimum is located at
$s_{m,\mathrm{inf}}\gg H_{\rm inf}/\kappa$ [see Eq.~\eqref{s_m_inf}],
and quantum fluctuations around it are estimated as
\begin{equation}
\langle (s-s_{m,\rm inf})^2\rangle \simeq \frac{3H_{\rm inf}^4}{8\pi^2m_{s}^2}
<\frac{3}{16\pi^2c}H_{\rm inf}^2 \ll 10^{-3}H_{\rm inf}^2,
\end{equation}
where
\begin{align}
m_s^2 &\equiv \frac{\partial^2V}{\partial s^2}(s_{m,\mathrm{inf}},0) \nonumber\\
&= 2cH_{\rm inf}^2 + \frac{1}{4}\left(c'^2+c'\sqrt{c'^2+8c}\right)H_{\rm inf}^2,
\end{align}
and it is assumed that only the mass term is important.
Thus, fluctuations along the radial direction are much smaller than the
expectation value $s_{m,\rm inf}$, and we may study fluctuations along
the angular direction by setting $s=s_{m,\rm inf}$.

The potential for the angular scalar field defined by $\chi \equiv
s_{m,\rm inf}\theta$ is given by
\begin{align}
V(\chi)&=-\frac{\kappa c'}{3\sqrt{2}}H_{\rm inf}s_{m,\rm inf}^3
\cos \lmk 3\frac{\chi}{s_{m,\rm inf}}\rmk  \nonumber\\
&=\frac{3\kappa c'}{2\sqrt{2}}H_{\rm inf}s_{m,\rm inf}\chi^2+..., 
\label{chipotential}
\end{align}
where the latter expression applies for small $\theta$, and we can
read off the mass of $\chi$ as
\begin{equation}
m^2_\chi=\frac{3c'}{4}\lmk c'+\sqrt{c'^2+8c}\rmk H_{\rm inf}^2.
\end{equation}

Let us estimate the number of domains where fluctuation along the angular
direction exceeds $|\theta|=\pi/3$ to make a domain wall somewhere within our
observable Universe using the approximated
potential [Eq.~\eqref{chipotential}] based on the peak theory of random
Gaussian fields [Ref.~\cite{Bardeen:1985tr}]. The desired quantity can be
calculated from the correlation function
\begin{equation}
C(|\vect z - \vect z'|)\equiv \langle \chi(\vect z)\chi(\vect z') \rangle
=\frac{3H^4_{\rm inf}}{8\pi^2 m_\chi^2}\lmk H_{\rm inf}|\vect z -\vect
z'| \rmk^{-\frac{2m_\chi^2}{3H^2_{\rm inf}}}, \label{correlation}
\end{equation}
and its derivatives at zero lag, such as
\begin{align}
\sigma_0^2&\equiv \langle \chi^2(0)\rangle=C(0)
=\frac{3H^4_{\rm inf}}{8\pi^2 m_\chi^2},  \\
\sigma_1^2&\equiv 3\left\langle \frac{\partial\chi}{\partial z_x}(\vect
 z)\frac{\partial\chi}{\partial z'_x}(\vect z')\right\rangle
= -  3\left.\frac{C'(r)}{r}\right|_{r\rightarrow {H^{-1}_{\rm inf}}}
=\frac{3H^4_{\rm inf}}{4\pi^2}, \\
\sigma_2^2&\equiv 15\left\langle 
\frac{\partial^2\chi}{\partial z_x^2}(\vect z)
\frac{\partial^2\chi}{\partial z'^2_y}(\vect z')\right\rangle
=15\left.\lmk
 \frac{C''(r)}{r^2}-\frac{C'(r)}{r^3}\rmk\right|_{r\rightarrow
 {H^{-1}_{\rm inf}}}\non\\
&=\frac{5H^4_{\rm inf}}{2\pi^2}(m_\chi^2+3H^2_{\rm inf}).
 \end{align}
Note that Eq.\ (\ref{correlation}) is applicable for $r \gtrsim H_{\rm
inf}^{-1}$, but it also reproduces the variance with zero lag, taking
$r=H_{\rm inf}^{-1}$.  Hence, we take  $r\rightarrow
{H^{-1}_{\rm inf}}$
when we consider the zero-lag limit in stochastic inflation [Ref.~\cite{Starobinsky:1994bd}].

Then according to Ref.~\cite{Bardeen:1985tr}, the number density of the
$\nu\sigma$ peak is calculated as
\begin{equation}
 n_{p}(\nu)d\nu =\frac{e^{-\frac{\nu^2}{2}}}{(2\pi)^2R_\ast^3}
G(\gamma,\gamma\nu)d\nu\equiv e^{-f(\nu)}d\nu,
\end{equation}
where $G(\gamma,w)$ is a function whose approximate form can be found
in Eq.\ (4.4) of Ref.~\cite{Bardeen:1985tr}.  Here $\gamma$ and $R_\ast$
are, respectively, given by
\begin{equation}
\gamma=\lmk\frac{6m_\chi^2}{10m_\chi^2+30H_{\rm
 inf}^2}\rmk^{\frac{1}{2}},~~
R_\ast=\frac{3}{(10m_\chi^2+30H_{\rm inf}^2)^{1/2}}.
\end{equation}
Assuming that the number of $e$-folds of inflation 
required to solve the horizon problem
is equal to 60, the number of domains with $|\theta|>\pi/3$ in the
observable Universe today is given by \cite{Yokoyama:1998xd}
\begin{align}
2N(>\nu_d)&=2e^{180}H^{-3}_{\rm inf}\int_{\nu_d}^{\infty}n_{p}(\nu)d\nu
\simeq \frac{2e^{180-f(\nu_d)}}{H^{3}_{\rm inf}f'(\nu_d)},\non \\
&\simeq\frac{e^{180-\nu_d^2/2}}{2\pi^2\nu_d}\lmk\frac{10m_\chi^2+30H^{2}_{\rm
 inf}}{9H^{2}_{\rm inf}}\rmk^{\frac{3}{2}}G(\gamma,\gamma\nu_d),
\end{align}
where $\nu_d\equiv \pi s_{m,\rm inf}/(3\sqrt{C(0)})$.
Requiring that the above expression should be smaller than unity, we find
only a mild constraint on $\kappa$. 
This can be easily seen from the exponent $-\nu_d^2/2=-\pi^4c'(c'+(c'^2+8c)^{1/2})^3/72\kappa^2$,
which gives a huge negative contribution for $\kappa\ll c'^2$.

Therefore to summarize, as long as the condition for the tracking behavior is satisfied,
quantum fluctuations during inflation 
are suppressed both along radial and angular
directions, so that the domain wall formation is always avoided.

\section{Thermal effects on $S$ after inflation and constraints\label{sec6}}

Finally we consider possible thermal effects on  the $S$ field after inflation.
As the amplitude $|S|$ decreases with time, interactions with light fields in
the thermal bath gradually come into play, which can destroy the $S$
field condensate to produce domain walls.  In particular, we must ensure
the following two conditions: 

\begin{enumerate}

\item Thermal  corrections of the form 
$\sim\lambda^2 T^2|S|^2$ should be small enough to satisfy $\lambda^2 T^2\ll \tilde{c}H^2$ 
in order not to affect the tracking evolution of
the $S$ field.

\item The amplitude of thermal fluctuation 
$\delta S(x)\sim T$ of the $S$ field should be well below that of the
background field $s_{m}$; otherwise they may cause transitions
into different domains to create domain walls.

\end{enumerate}

Since $s_m\sim \sqrt{\tilde c}H/\kappa$ and
$\lambda\simeq\kappa$, we see that these two requirements lead to 
the identical condition on the model parameters, which is solely determined by
the dynamics of later times, since $H$ decreases faster
than $T$ after inflation.  
Hence, we have only to consider the constraint at the time when
$S$ has almost reached the global minimum, i.e., ${\tilde c}H^2\sim m_S^2$.
Therefore, we may impose the following condition to avoid destruction of the
condensate\footnote{To confirm this observation, we have
numerically solved the field
equation $\ddot{S}+3H\dot{S}+\partial V/\partial S=0$ with the potential
given by $V\simeq -{\tilde c}H^2|S|^2+\kappa^2|S|^4 + m_S^2|S|^2 + (\kappa
A_{\kappa}S^3/3-\kappa\tilde{c'}HS^3/3 + \mathrm{h.c.})$, which gives a good approximation
around ${\tilde c}H^2\gtrsim m_S^2$. We have
confirmed that for reasonable values of
the model parameters the $S$ field does not move to other minima if
the deviation from $s_m/\sqrt{2}$ is less than $\mathcal{O}(10)\%$, which
we use as the criterion shown in Eq.~\eqref{criterion}.}:
\begin{equation}
\frac{\lambda T}{\sqrt{\tilde c}H} < 0.1\quad \mathrm{at} \quad 
{\tilde c}H^2=m_S^2. \label{criterion}
\end{equation}
Let us quantify Eq.~\eqref{criterion} more explicitly by
considering the evolution of the Universe after inflation.  Just after
inflation, the inflaton starts to oscillate around the minimum of
its own potential, and it eventually decays into the (MS)SM degrees of
freedom to reheat the Universe. There are two distinct possibilities:
(1) the $S$ field reaches its global minimum during the 
inflaton-oscillation dominated phase before reheating is completed, 
and (2) the $S$ field reaches its global minimum in the radiation dominated epoch after reheating.

In the first scenario, the temperature of the thermal bath is given by
$T\simeq (HT_R^2M_{\rm Pl})^{1/4}$~\cite{Kolb:1990vq}, where $T_R$ is the
reheating temperature. At the relevant epoch we find $T\simeq
{\tilde c}^{-1/8}(m_ST_R^2M_{\rm Pl})^{1/4}$, and
Eq.~\eqref{criterion} leads to the following bound:
\begin{equation}
T_R < 2\times 10^3\mathrm{GeV}\ {\tilde c}^{\frac{1}{4}}
\left(\frac{\lambda}{10^{-5}}\right)^{-2}\left(\frac{m_S}{1\mathrm{TeV}}\right)^{\frac{3}{2}}.
 \label{constraint1}
\end{equation}
Since $T\simeq {\tilde c}^{-1/8}(m_ST_R^2M_{\rm Pl})^{1/4}>T_R$
by assumption, we obtain an upper bound on $T_R$:
\begin{equation}
T_R < 5\times 10^{10}\mathrm{GeV}\ {\tilde c}^{-\frac{1}{4}}\left(\frac{m_S}{1\mathrm{TeV}}\right)^{\frac{1}{2}}. \label{constraint2}
\end{equation}

If the condition given by Eq.~\eqref{constraint2} is not satisfied, the
$S$ field reaches its global minimum in the radiation dominated epoch
after reheating. From the relation $H\sim T^2/M_{\rm Pl}$, 
we have $T\simeq \tilde{c}^{-1/4}(m_SM_{\rm Pl})^{1/2}$ 
at ${\tilde c}H^2=m_S^2$,
and by using Eq.~\eqref{criterion}, we obtain
\begin{equation}
\lambda < 2\times 10^{-9}\ {\tilde c}^{\frac{1}{4}}\left(\frac{m_S}{1\mathrm{TeV}}\right)^{\frac{1}{2}}. \label{constraint3}
\end{equation}
Note that the above condition makes sense only if there exists a
large negative Hubble-induced mass term $-{\tilde c}H^2|S|^2$ and $A$ terms
$-(\kappa\tilde{c'}HS^3+\mathrm{h.c.})$ even after the reheating is completed.
The existence of the negative Hubble-induced mass term
is foreseeable due to the SUSY breaking contributions from the thermal bath~\cite{Lyth:2004nx,Kawasaki:2011zi}.
The phase of the coefficient $\tilde{c'}$ of the Hubble-induced $A$ term can change after reheating
since the K\"ahler structures might change at that epoch, which induces a rotation of the $S$ field in the angular direction.
This fact does not spoil the tracking mechanism as long as the magnitudes of the coefficients remain sufficiently large.
The reason may be understood as follows: 
the angular motion induced by the change of the K\"ahler structures proceeds adiabatically with a Hubble time scale,
since the transition to the radiation dominated phase happens smoothly at that epoch.
For such a smooth transition, the deviation from the minimum of the effective potential induced by the angular motion
is at most $\mathcal{O}(\tilde{c}^{-1},\tilde{c'}^{-2})$, which remains negligible for large coefficients.

%\if0

\begin{figure}[htbp]
\includegraphics[width=0.45\textwidth]{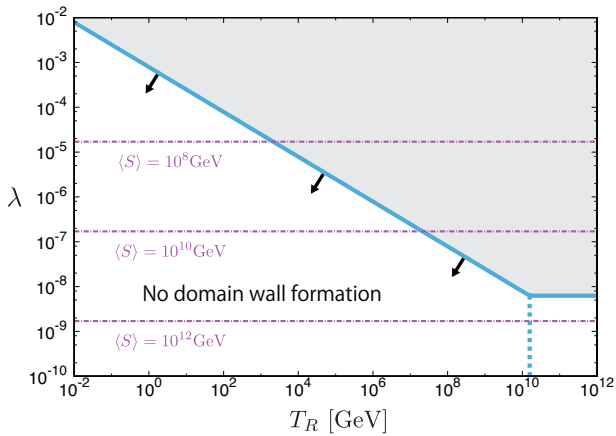}
\caption{Conditions to avoid the domain wall formation on the parameter space of $(T_R,\lambda)$.
Domain walls are not produced in the region below the blue line corresponding to Eqs.~\eqref{constraint1} and~\eqref{constraint3}.
The region to the right side of the blue dotted line corresponding to Eq.~\eqref{constraint2} also leads to the domain wall formation
if the large negative Hubble-induced mass is absent in the radiation dominated epoch after inflation.
We also plot the values of the VEV $\langle S\rangle$ at the global minimum [Eq.~\eqref{S_VEV_global}] as purple chain lines.
Here we fix the values of other parameters as ${\tilde c}=100$, $m_S=1\mathrm{TeV}$ and $|A_{\kappa}|=4\mathrm{TeV}$.}
\label{fig2}
\end{figure}

%\fi

Figure~\ref{fig2} summarizes our analyses.
Combining two cases described above, we show the allowed parameter range  
for $T_R$ and $\lambda (\simeq \kappa)$.
From the plot we see that the resolution of the domain wall problem requires a
small value for the Higgs-singlet coupling besides the large enough
Hubble-induced mass. 

\section{Cosmological consequences for electroweak baryogenesis and dark matter\label{sec7}}

The results obtained in this paper lead to several
consequences for cosmology in the NMSSM.  First, for sufficiently small
couplings the singlet sector would decouple from the thermal bath.  A
dominant interaction process between the singlet fields and the MSSM
fields is the $2 \to 2$ interaction involving Higgsinos and weak bosons
(and their SUSY partners), whose rate is roughly estimated as
$\Gamma\sim\alpha\lambda^2 T$, where $\alpha$ is the weak coupling
strength. Since this interaction rate is exponentially suppressed after
the temperature of the thermal bath becomes less than
$m_S\sim\mathcal{O}(\mathrm{TeV})$, and the Hubble parameter $H \propto
T^2$ decays faster than $\Gamma$, we expect that the fields in the
singlet sector would never thermalize if the condition $\Gamma<H$ is satisfied
at $T\sim m_S$.  This occurs if the value of the coupling satisfies
\begin{equation}
\lambda < \alpha^{-\frac{1}{2}}\left(\frac{m_S}{M_{\rm Pl}}\right)^{\frac{1}{2}}\sim
\mathcal{O}(10^{-7}\mathchar`-10^{-6})\,.
\end{equation}
For such a small
coupling the singlino state cannot be produced from the thermal bath,
and it would not contribute to the dark matter abundance.  Finally, the
$S$ field falls in the global minimum much earlier than the epoch of the
electroweak phase transition $T\sim\mathcal{O}(100)\mathrm{GeV}$, if the
tracking mechanism works. Therefore, the first order phase transition is
not likely to be realized in this case.
These facts would have an important consequence for 
the electroweak baryogenesis within NMSSM and the mass range for the dark matter,
which would now primarily contain the Higgsino component~\cite{Balazs:2013cia}.

\section{Conclusion and discussions\label{sec8}}

In this paper, we have shown that the domain wall problem of the NMSSM can be resolved via the tracking mechanism if the following conditions are satisfied:
(1)~There exist negative Hubble-induced corrections with sufficiently large coefficients [see Eq.~\eqref{bound_c}]
in the effective potential for the $S$ field during and after inflation.
(2)~The $A_{\kappa}$ term is slightly larger than the soft mass $m_S$ [see Eq.~\eqref{condition_A_term}]
in order to prevent the $S$ field from rotating in the phase direction at later times.
(3)~The thermal effects must remain irrelevant; see Eq.~\eqref{criterion}. This requirement leads to the constraints on 
the coupling strength and the reheating temperature as shown in Fig.~\ref{fig2}. Our study has many important phenomenological consequences 
for LHC and dark matter creation and detection within NMSSM.

Let us briefly mention a particular relevance for the
future experimental studies of NMSSM.  Typically, in NMSSM the Higgs sector 
is enlarged due to the presence of a singlet,  and for reasonably large couplings, i.e., $\lambda,~\kappa\sim 0.1$, 
Higgs-to-Higgs decays can possibly be observed in the forthcoming LHC
experiment~\cite{Christensen:2013dra}.
According to Fig.~\ref{fig2}, such a discovery at a large coupling regime $\lambda
\sim \mathcal{O}(0.1)$ indicates the formation of domain walls; then we
must seriously take into account their cosmological evolution~\cite{Abel:1995wk,Kadota:2015dza}.  

Another important LHC signature would be obtained in the small coupling regime, where the
singlet states decouple from the MSSM sector.  In this case, we would expect
to observe displaced vertices from the long-lived next-to-LSP~\cite{Ellwanger:1997jj}, which can also be 
helpful to falsify our scenario.

%%%%%%%%%%%%%%%%%%%%%%%%%%%%%%%%%%%%%%%%%%%%%%%%%%%%%%%%%%%%%%%%%%%%%%
%%%%%%%%%%%%%%%%%%%%%%%%%%%%%%%%%%%%%%%%%%%%%%%%%%%%%%%%%%%%%%%%%%%%%%

\begin{acknowledgments}
A.~M.~thanks Csaba Balasz for a discussion about some aspects of
NMSSM. K.~S.~thanks Koichi Hamaguchi, Kyohei Mukaida, Kazunori
Nakayama, and Osamu Seto for a discussion about the stability of the tracking behavior
of the singlet field. A.~M.~and M.~Y.~thank the Japan Society for the Promotion
of Science (JSPS) Invitation
Fellowship for Research in Japan. A.~M.~is supported by the STFC Grant
ST/J000418/1. K.~S.~is supported by the JSPS through research fellowships. This work was in part supported
by the JSPS Grant-in-Aid for Scientific Research Nos. 25287054 (M.Y.),
26610062 (M.Y.), 15H02082 (J.Y.), and the JSPS Grant-in-Aid for Scientific
Research on Innovative Areas No. 15H05888 (M.Y. and J.Y.).
\end{acknowledgments}

%%%%%%%%%%%%%%%%%%%%%%%%%%%%%%%%%%%%%%%%%%%%%%%%%%%%%%%%%%%%%%%%%%%%%%
%%%%%%%%%%%%%%%%%%%%%%%%%%%%%%%%%%%%%%%%%%%%%%%%%%%%%%%%%%%%%%%%%%%%%%


\begin{thebibliography}{44}
\expandafter\ifx\csname natexlab\endcsname\relax\def\natexlab#1{#1}\fi
\expandafter\ifx\csname bibnamefont\endcsname\relax
  \def\bibnamefont#1{#1}\fi
\expandafter\ifx\csname bibfnamefont\endcsname\relax
  \def\bibfnamefont#1{#1}\fi
\expandafter\ifx\csname citenamefont\endcsname\relax
  \def\citenamefont#1{#1}\fi
\expandafter\ifx\csname url\endcsname\relax
  \def\url#1{\texttt{#1}}\fi
\expandafter\ifx\csname urlprefix\endcsname\relax\def\urlprefix{URL }\fi
\providecommand{\bibinfo}[2]{#2}
\providecommand{\eprint}[2][]{\url{#2}}

\bibitem[{\citenamefont{Nilles}(1984)}]{Nilles:1983ge}
\bibinfo{author}{\bibfnamefont{H.~P.} \bibnamefont{Nilles}},
  \bibinfo{journal}{Phys.Rept.} \textbf{\bibinfo{volume}{110}},
  \bibinfo{pages}{1} (\bibinfo{year}{1984}).

\bibitem[{\citenamefont{Martin}(2010)}]{Martin:1997ns}
\bibinfo{author}{\bibfnamefont{S.~P.} \bibnamefont{Martin}},
  \bibinfo{journal}{Adv.Ser.Direct.High Energy Phys.}
  \textbf{\bibinfo{volume}{21}}, \bibinfo{pages}{1} (\bibinfo{year}{2010})
[hep-ph/9709356].
\bibitem[{\citenamefont{Enqvist and Mazumdar}(2003)}]{Enqvist:2003gh}
\bibinfo{author}{\bibfnamefont{K.}~\bibnamefont{Enqvist}} \bibnamefont{and}
  \bibinfo{author}{\bibfnamefont{A.}~\bibnamefont{Mazumdar}},
  \bibinfo{journal}{Phys. Rept.} \textbf{\bibinfo{volume}{380}},
  \bibinfo{pages}{99} (\bibinfo{year}{2003})
  [hep-ph/0209244].
\bibitem[{\citenamefont{Jungman et~al.}(1996)\citenamefont{Jungman,
  Kamionkowski, and Griest}}]{Jungman:1995df}
\bibinfo{author}{\bibfnamefont{G.}~\bibnamefont{Jungman}},
  \bibinfo{author}{\bibfnamefont{M.}~\bibnamefont{Kamionkowski}},
  \bibnamefont{and} \bibinfo{author}{\bibfnamefont{K.}~\bibnamefont{Griest}},
  \bibinfo{journal}{Phys.Rept.} \textbf{\bibinfo{volume}{267}},
  \bibinfo{pages}{195} (\bibinfo{year}{1996})
  [hep-ph/9506380].

\bibitem[{\citenamefont{Lyth and Riotto}(1999)}]{Lyth:1998xn}
\bibinfo{author}{\bibfnamefont{D.~H.} \bibnamefont{Lyth}} \bibnamefont{and}
  \bibinfo{author}{\bibfnamefont{A.}~\bibnamefont{Riotto}},
  \bibinfo{journal}{Phys. Rept.} \textbf{\bibinfo{volume}{314}},
  \bibinfo{pages}{1} (\bibinfo{year}{1999})
  [hep-ph/9807278].

\bibitem[{\citenamefont{Mazumdar and Rocher}(2011)}]{Mazumdar:2010sa}
\bibinfo{author}{\bibfnamefont{A.}~\bibnamefont{Mazumdar}} \bibnamefont{and}
  \bibinfo{author}{\bibfnamefont{J.}~\bibnamefont{Rocher}},
  \bibinfo{journal}{Phys. Rept.} \textbf{\bibinfo{volume}{497}},
  \bibinfo{pages}{85} (\bibinfo{year}{2011})
  [arXiv:1001.0993 [hep-ph]].
  
\bibitem{Yamaguchi:2011kg} 
  M.~Yamaguchi,
  %``Supergravity based inflation models: a review,''
  Class.\ Quant.\ Grav.\  {\bf 28}, 103001 (2011)
  [arXiv:1101.2488 [astro-ph.CO]].
  
\bibitem{Murayama:1992ua} 
  H.~Murayama, H.~Suzuki, T.~Yanagida and J.~Yokoyama,
  %``Chaotic inflation and baryogenesis by right-handed sneutrinos,''
  Phys.\ Rev.\ Lett.\  {\bf 70}, 1912 (1993).
  
\bibitem{Murayama:1993xu} 
  H.~Murayama, H.~Suzuki, T.~Yanagida and J.~Yokoyama,
  %``Chaotic inflation and baryogenesis in supergravity,''
  Phys.\ Rev.\ D {\bf 50}, R2356 (1994)
  [hep-ph/9311326].
  
\bibitem{Kadota:2005mt} 
  K.~Kadota and J.~Yokoyama,
  %``D-term inflation and leptogenesis by right-handed sneutrino,''
  Phys.\ Rev.\ D {\bf 73}, 043507 (2006)
  [hep-ph/0512221].

\bibitem[{\citenamefont{Allahverdi et~al.}(2006)\citenamefont{Allahverdi,
  Enqvist, Garcia-Bellido, and Mazumdar}}]{Allahverdi:2006iq}
\bibinfo{author}{\bibfnamefont{R.}~\bibnamefont{Allahverdi}},
  \bibinfo{author}{\bibfnamefont{K.}~\bibnamefont{Enqvist}},
  \bibinfo{author}{\bibfnamefont{J.}~\bibnamefont{Garcia-Bellido}},
  \bibnamefont{and} \bibinfo{author}{\bibfnamefont{A.}~\bibnamefont{Mazumdar}},
  \bibinfo{journal}{Phys. Rev. Lett.} \textbf{\bibinfo{volume}{97}},
  \bibinfo{pages}{191304} (\bibinfo{year}{2006})
  [hep-ph/0605035].

\bibitem[{\citenamefont{Allahverdi et~al.}(2007)\citenamefont{Allahverdi,
  Enqvist, Garcia-Bellido, Jokinen, and Mazumdar}}]{Allahverdi:2006we}
\bibinfo{author}{\bibfnamefont{R.}~\bibnamefont{Allahverdi}},
  \bibinfo{author}{\bibfnamefont{K.}~\bibnamefont{Enqvist}},
  \bibinfo{author}{\bibfnamefont{J.}~\bibnamefont{Garcia-Bellido}},
  \bibinfo{author}{\bibfnamefont{A.}~\bibnamefont{Jokinen}}, \bibnamefont{and}
  \bibinfo{author}{\bibfnamefont{A.}~\bibnamefont{Mazumdar}},
  \bibinfo{journal}{JCAP} \textbf{\bibinfo{volume}{0706}}, \bibinfo{pages}{019}
  (\bibinfo{year}{2007})
  [hep-ph/0610134].
  
  \bibitem{Kamada:2009hy} 
  K.~Kamada and J.~Yokoyama,
  %``On the realization of the MSSM inflation,''
  Prog.\ Theor.\ Phys.\  {\bf 122}, 969 (2009)
  [arXiv:0906.3402 [hep-ph]].
  
  \bibitem{Ferrara:2010yw} 
  S.~Ferrara, R.~Kallosh, A.~Linde, A.~Marrani and A.~Van Proeyen,
  %``Jordan Frame Supergravity and Inflation in NMSSM,''
  Phys.\ Rev.\ D {\bf 82}, 045003 (2010)
  [arXiv:1004.0712 [hep-th]].  
  
  \bibitem{Ferrara:2010in} 
  S.~Ferrara, R.~Kallosh, A.~Linde, A.~Marrani and A.~Van Proeyen,
  %``Superconformal Symmetry, NMSSM, and Inflation,''
  Phys.\ Rev.\ D {\bf 83}, 025008 (2011)
  [arXiv:1008.2942 [hep-th]].
  
\bibitem{Affleck:1984fy} 
  I.~Affleck and M.~Dine,
  %``A New Mechanism for Baryogenesis,''
  Nucl.\ Phys.\ B {\bf 249}, 361 (1985).

  \bibitem{Nilles-1}
  H.P. Nilles, M. Srednicki and D. Wyler, Phys. Lett. B 120, 346 (1983); J. Fr\`ere, D.R.T. Jones and S. Raby, Nucl. Phys. B 222, 11 (1983); J.P. Derendinger and C.A. Savoy, Nucl. Phys. B 237, 307 (1984).

\bibitem[{\citenamefont{Kim and Nilles}(1984)}]{Kim:1983dt}
\bibinfo{author}{\bibfnamefont{J.~E.} \bibnamefont{Kim}} \bibnamefont{and}
  \bibinfo{author}{\bibfnamefont{H.~P.} \bibnamefont{Nilles}},
  \bibinfo{journal}{Phys.Lett.} \textbf{\bibinfo{volume}{B138}},
  \bibinfo{pages}{150} (\bibinfo{year}{1984}).

\bibitem[{\citenamefont{Ellwanger et~al.}(2010)\citenamefont{Ellwanger,
  Hugonie, and Teixeira}}]{Ellwanger:2009dp}
\bibinfo{author}{\bibfnamefont{U.}~\bibnamefont{Ellwanger}},
  \bibinfo{author}{\bibfnamefont{C.}~\bibnamefont{Hugonie}}, \bibnamefont{and}
  \bibinfo{author}{\bibfnamefont{A.~M.} \bibnamefont{Teixeira}},
  \bibinfo{journal}{Phys.Rept.} \textbf{\bibinfo{volume}{496}},
  \bibinfo{pages}{1} (\bibinfo{year}{2010})
  [arXiv:0910.1785 [hep-ph]].

  \bibitem{Frieman:1991tu} 
  J.~A.~Frieman, C.~T.~Hill and R.~Watkins,
  %``Late time cosmological phase transitions. 1. Particle physics models and cosmic evolution,''
  Phys.\ Rev.\ D {\bf 46}, 1226 (1992).
  
  \bibitem{Brandenberger:2003ge} 
  R.~Brandenberger, D.~A.~Easson and A.~Mazumdar,
  %``Inflation and brane gases,''
  Phys.\ Rev.\ D {\bf 69}, 083502 (2004)
  [hep-th/0307043].
  %%CITATION = HEP-TH/0307043;%%

\bibitem[{\citenamefont{Zeldovich et~al.}(1974)\citenamefont{Zeldovich,
  Kobzarev, and Okun}}]{Zeldovich:1974uw}
\bibinfo{author}{\bibfnamefont{Y.}~\bibnamefont{Zeldovich}},
  \bibinfo{author}{\bibfnamefont{I.~Y.} \bibnamefont{Kobzarev}},
  \bibnamefont{and} \bibinfo{author}{\bibfnamefont{L.}~\bibnamefont{Okun}},
  \bibinfo{journal}{Zh.Eksp.Teor.Fiz.} \textbf{\bibinfo{volume}{67}},
  \bibinfo{pages}{3} (\bibinfo{year}{1974}).

\bibitem[{\citenamefont{Panagiotakopoulos and
  Tamvakis}(1999)}]{Panagiotakopoulos:1998yw}
\bibinfo{author}{\bibfnamefont{C.}~\bibnamefont{Panagiotakopoulos}}
  \bibnamefont{and} \bibinfo{author}{\bibfnamefont{K.}~\bibnamefont{Tamvakis}},
  \bibinfo{journal}{Phys.Lett.} \textbf{\bibinfo{volume}{B446}},
  \bibinfo{pages}{224} (\bibinfo{year}{1999})
  [hep-ph/9809475].
  
 \bibitem{McDonald:1997vy} 
  J.~McDonald,
  %``Spontaneous discrete symmetry breaking during inflation and the NMSSM domain wall problem,''
  Nucl.\ Phys.\ B {\bf 530}, 325 (1998)
  [hep-ph/9709512].  
  
 \bibitem[{\citenamefont{Ellwanger et~al.}(1997)\citenamefont{Ellwanger,
  Rausch~de Traubenberg, and Savoy}}]{Ellwanger:1996gw}
\bibinfo{author}{\bibfnamefont{U.}~\bibnamefont{Ellwanger}},
  \bibinfo{author}{\bibfnamefont{M.}~\bibnamefont{Rausch~de Traubenberg}},
  \bibnamefont{and} \bibinfo{author}{\bibfnamefont{C.~A.} \bibnamefont{Savoy}},
  \bibinfo{journal}{Nucl.Phys.} \textbf{\bibinfo{volume}{B492}},
  \bibinfo{pages}{21} (\bibinfo{year}{1997}) 
  [hep-ph/9611251].  
  
\bibitem[{\citenamefont{Vilenkin and Ford}(1982)}]{Vilenkin:1982wt}
\bibinfo{author}{\bibfnamefont{A.}~\bibnamefont{Vilenkin}} \bibnamefont{and}
  \bibinfo{author}{\bibfnamefont{L.~H.} \bibnamefont{Ford}},
  \bibinfo{journal}{Phys. Rev.} \textbf{\bibinfo{volume}{D26}},
  \bibinfo{pages}{1231} (\bibinfo{year}{1982}).

\bibitem[{\citenamefont{Linde}(1982)}]{Linde:1982uu}
\bibinfo{author}{\bibfnamefont{A.~D.} \bibnamefont{Linde}},
  \bibinfo{journal}{Phys. Lett.} \textbf{\bibinfo{volume}{B116}},
  \bibinfo{pages}{335} (\bibinfo{year}{1982}).

\bibitem[{\citenamefont{Starobinsky}(1982)}]{Starobinsky:1982ee}
\bibinfo{author}{\bibfnamefont{A.~A.} \bibnamefont{Starobinsky}},
  \bibinfo{journal}{Phys. Lett.} \textbf{\bibinfo{volume}{B117}},
  \bibinfo{pages}{175} (\bibinfo{year}{1982}).
  
\bibitem{Starobinsky:1994bd} 
  A.~A.~Starobinsky and J.~Yokoyama,
  %``Equilibrium state of a selfinteracting scalar field in the De Sitter background,''
  Phys.\ Rev.\ D {\bf 50}, 6357 (1994)
  [astro-ph/9407016].

\bibitem[{\citenamefont{Kofman et~al.}(1994)\citenamefont{Kofman, Linde, and
  Starobinsky}}]{Kofman:1994rk}
\bibinfo{author}{\bibfnamefont{L.}~\bibnamefont{Kofman}},
  \bibinfo{author}{\bibfnamefont{A.~D.} \bibnamefont{Linde}}, \bibnamefont{and}
  \bibinfo{author}{\bibfnamefont{A.~A.} \bibnamefont{Starobinsky}},
  \bibinfo{journal}{Phys. Rev. Lett.} \textbf{\bibinfo{volume}{73}},
  \bibinfo{pages}{3195} (\bibinfo{year}{1994})
  [hep-th/9405187].
  
 \bibitem{Sato:2015dga} 
  K.~Sato and J.~Yokoyama,
  %``Inflationary cosmology: First 30+ years,''
  Int.\ J.\ Mod.\ Phys.\ D {\bf 24}, no. 11, 1530025 (2015).
  %%CITATION = IMPAE,D24,1530025;%%  
  
\bibitem[{\citenamefont{Dine et~al.}(1995)\citenamefont{Dine, Randall, and
  Thomas}}]{Dine:1995uk}
\bibinfo{author}{\bibfnamefont{M.}~\bibnamefont{Dine}},
  \bibinfo{author}{\bibfnamefont{L.}~\bibnamefont{Randall}}, \bibnamefont{and}
  \bibinfo{author}{\bibfnamefont{S.~D.} \bibnamefont{Thomas}},
  \bibinfo{journal}{Phys. Rev. Lett.} \textbf{\bibinfo{volume}{75}},
  \bibinfo{pages}{398} (\bibinfo{year}{1995})
  [hep-ph/9503303];
\bibinfo{author}{\bibfnamefont{M.}~\bibnamefont{Dine}},
  \bibinfo{author}{\bibfnamefont{L.}~\bibnamefont{Randall}}, \bibnamefont{and}
  \bibinfo{author}{\bibfnamefont{S.~D.} \bibnamefont{Thomas}},
  \bibinfo{journal}{Nucl. Phys.} \textbf{\bibinfo{volume}{B458}},
  \bibinfo{pages}{291} (\bibinfo{year}{1996})
  [hep-ph/9507453].
  
\bibitem[{\citenamefont{Choudhury et~al.}(2014)\citenamefont{Choudhury,
  Mazumdar, and Pukartas}}]{Choudhury:2014sxa}
\bibinfo{author}{\bibfnamefont{S.}~\bibnamefont{Choudhury}},
  \bibinfo{author}{\bibfnamefont{A.}~\bibnamefont{Mazumdar}}, \bibnamefont{and}
  \bibinfo{author}{\bibfnamefont{E.}~\bibnamefont{Pukartas}},
  \bibinfo{journal}{JHEP} \textbf{\bibinfo{volume}{04}}, \bibinfo{pages}{077}
  (\bibinfo{year}{2014})
  [arXiv:1402.1227 [hep-th]].
  
  \bibitem{Linde:1996cx} 
  A.~D.~Linde,
  %``Relaxing the cosmological moduli problem,''
  Phys.\ Rev.\ D {\bf 53}, R4129 (1996)
  [hep-th/9601083].
  
\bibitem{Bardeen:1985tr} 
  J.~M.~Bardeen, J.~R.~Bond, N.~Kaiser and A.~S.~Szalay,
  %``The Statistics of Peaks of Gaussian Random Fields,''
  Astrophys.\ J.\  {\bf 304}, 15 (1986).
  
\bibitem{Yokoyama:1998xd} 
  J.~Yokoyama,
  %``Cosmological constraints on primordial black holes produced in the near critical gravitational collapse,''
  Phys.\ Rev.\ D {\bf 58}, 107502 (1998)
  [gr-qc/9804041].
  
  \bibitem{Kolb:1990vq} 
  E.~W.~Kolb and M.~S.~Turner,
  %``The Early Universe,''
  Front.\ Phys.\  {\bf 69}, 1 (1990).
  %%CITATION = FRPHA,69,1;%%
  
\bibitem[{\citenamefont{Kawasaki and Takesako}(2012)}]{Kawasaki:2011zi}
\bibinfo{author}{\bibfnamefont{M.}~\bibnamefont{Kawasaki}} \bibnamefont{and}
  \bibinfo{author}{\bibfnamefont{T.}~\bibnamefont{Takesako}},
  \bibinfo{journal}{Phys. Lett.} \textbf{\bibinfo{volume}{B711}},
  \bibinfo{pages}{173} (\bibinfo{year}{2012})
  [arXiv:1112.5823 [hep-ph]].

\bibitem[{\citenamefont{Lyth and Moroi}(2004)}]{Lyth:2004nx}
\bibinfo{author}{\bibfnamefont{D.~H.} \bibnamefont{Lyth}} \bibnamefont{and}
  \bibinfo{author}{\bibfnamefont{T.}~\bibnamefont{Moroi}},
  \bibinfo{journal}{JHEP} \textbf{\bibinfo{volume}{05}}, \bibinfo{pages}{004}
  (\bibinfo{year}{2004})
  [hep-ph/0402174].

\bibitem{Balazs:2013cia} 
  C.~Balazs, A.~Mazumdar, E.~Pukartas and G.~White,
  %``Baryogenesis, dark matter and inflation in the Next-to-Minimal Supersymmetric Standard Model,''
  JHEP {\bf 1401}, 073 (2014)
  [arXiv:1309.5091 [hep-ph]].
  %%CITATION = ARXIV:1309.5091;%%

\bibitem[{\citenamefont{Christensen et~al.}(2013)\citenamefont{Christensen,
  Han, Liu, and Su}}]{Christensen:2013dra}
\bibinfo{author}{\bibfnamefont{N.~D.} \bibnamefont{Christensen}},
  \bibinfo{author}{\bibfnamefont{T.}~\bibnamefont{Han}},
  \bibinfo{author}{\bibfnamefont{Z.}~\bibnamefont{Liu}}, \bibnamefont{and}
  \bibinfo{author}{\bibfnamefont{S.}~\bibnamefont{Su}}, \bibinfo{journal}{JHEP}
  \textbf{\bibinfo{volume}{1308}}, \bibinfo{pages}{019} (\bibinfo{year}{2013})
  [arXiv:1303.2113 [hep-ph]].
  
\bibitem[{\citenamefont{Abel et~al.}(1995)\citenamefont{Abel, Sarkar, and
  White}}]{Abel:1995wk}
\bibinfo{author}{\bibfnamefont{S.}~\bibnamefont{Abel}},
  \bibinfo{author}{\bibfnamefont{S.}~\bibnamefont{Sarkar}}, \bibnamefont{and}
  \bibinfo{author}{\bibfnamefont{P.}~\bibnamefont{White}},
  \bibinfo{journal}{Nucl.Phys.} \textbf{\bibinfo{volume}{B454}},
  \bibinfo{pages}{663} (\bibinfo{year}{1995})
  [hep-ph/9506359].

\bibitem[{\citenamefont{Kadota et~al.}(2015)\citenamefont{Kadota, Kawasaki, and
  Saikawa}}]{Kadota:2015dza}
\bibinfo{author}{\bibfnamefont{K.}~\bibnamefont{Kadota}},
  \bibinfo{author}{\bibfnamefont{M.}~\bibnamefont{Kawasaki}}, \bibnamefont{and}
  \bibinfo{author}{\bibfnamefont{K.}~\bibnamefont{Saikawa}},
   JCAP {\bf 1510}, 041 (2015)
  [arXiv:1503.06998 [hep-ph]].

\bibitem[{\citenamefont{Ellwanger and Hugonie}(1998)}]{Ellwanger:1997jj}
\bibinfo{author}{\bibfnamefont{U.}~\bibnamefont{Ellwanger}} \bibnamefont{and}
  \bibinfo{author}{\bibfnamefont{C.}~\bibnamefont{Hugonie}},
  \bibinfo{journal}{Eur. Phys. J.} \textbf{\bibinfo{volume}{C5}},
  \bibinfo{pages}{723} (\bibinfo{year}{1998})
  [hep-ph/9712300].

\end{thebibliography}
\end{document}